\documentclass[12pt]{article}
\def\ca{\c{c}\~{a}}

\newcommand{\eff}{\mathrm{eff}}

\usepackage{epsfig}

\begin{document}

{\centerline{\bf Extended NJL model with eight-quark interactions}}
\vspace{0.5cm}

{\centerline{A. A. Osipov$^{1,2}$, B. Hiller$^1$, A. H. Blin$^1$, J. Moreira$^1$}}
\vspace{0.5cm}

\noindent{\it{$^1$Centro de F\'{\i}sica Computacional, Departamento de
F\'{\i}sica da Universidade de Coimbra, 3004-516 Coimbra, Portugal}

\noindent{$^2$Dzhelepov Laboratory of Nuclear Problems, JINR 141980 Dubna,
Russia}}
\vspace{0.5cm}

\begin{abstract}
We present the results obtained in the three-flavour ($N_f=3$)
Nambu--Jona-Lasinio model which is extended by the $U(1)_A$ breaking
six-quark 't Hooft interaction and eight-quark interactions. We address
the problem of stability, and some phenomenological consequences of the
models with multi-quark interactions.
\end{abstract}

A number of instructive models in low-energy QCD assume the existence of
underlying multi-quark interactions and their importance for the physics of
hadrons. They are ef\mbox{}f\mbox{}icient in the description of spontaneous
chiral symmetry breaking ($\chi$SB), and in the study of the quark structure
of light mesons. The Nambu--Jona-Lasinio (NJL) model \cite{Nambu:1961} is a
well-known example, where the local chiral symmetric four-fermion interactions
under some conditions lead to the formation of fermion-antifermion bound
states and as a result describe the $\chi$SB phenomenon. A modif\mbox{}ied
form of these interactions has been widely considered to derive the QCD
ef\mbox{}fective action at large distances
\cite{Eguchi:1976}-\cite{Bijnens:1993}.

One might ask if higher order multi-quark interactions are of importance.
Indeed, along the lines suggested by an instanton-gas model, it can be argued
\cite{Simonov:1997} that there exists an inf\mbox{}inite set of multi-quark
terms in the ef\mbox{}fective quark Lagrangian starting from the NJL
four-quark piece. In particular, the famous 't Hooft determinantal
interaction \cite{Hooft:1978} automatically appears if one keeps only the
zero mode contribution in the mode expansion of the ef\mbox{}fective
Lagrangian. This $2N_f$ multi-quark term manifestly violates the $U_A(1)$
axial symmetry of the QCD Lagrangian, of\mbox{}fering a way out of the
$U_A(1)$ problem.

The structure of QCD-motivated models at low energies with ef\mbox{}fective
many-fermion interaction and a f\mbox{}inite cutof\mbox{}f in the
symmetry-breaking regime has been also considered in \cite{Andrianov:1993},
where the authors, using the $1/N_c$ arguments and the f\mbox{}ine-tuning
condition in providing the scale invariance, classif\mbox{}ied the set of
quasilocal vertices relevant for dynamical $\chi$SB. It has been found
this way that in such ef\mbox{}fective models the vertices with four, six
and eight fermions only should be retained in four-dimensional space-time.

Thus, it is tempting to consider the intuitive picture that describes
the QCD vacuum based on a series of multi-quark interactions
ref\mbox{}lecting several tractable features of QCD, which include aspects of
chiral symmetry and of the $1/N_c$ expansion. The bosonization of quark
degrees of freedom leads then to the desirable ef\mbox{}fective Lagrangian
with matter f\mbox{}ields and a stable chiral asymmetric vacuum.

The NJL-type model with the $U_A(1)$ axial symmetry breaking by the 't Hooft
determinant has been studied in the mean field approximation
\cite{Bernard:1988}-\cite{Hatsuda:1994} for a long time. Numerous
phenomenological applications show that the results of such an approach meet
the expectations. Nevertheless, in this picture there is an apparent problem:
the mean field potential is unbounded from below, and the 't Hooft term is
the direct source of such an instability. A consistent approach requires
obviously a stable hadronic vacuum.

To cure this disease of the model we consider the system of light
quarks $u, d, s$ with multi-fermion interactions described by the
Lagrangian
\begin{equation}
\label{efflag}
  {\cal L}_\eff =\bar{q}(i\gamma^\mu\partial_\mu - m)q
          +{\cal L}_{4q} + {\cal L}_{6q}
          +{\cal L}_{8q}.
\end{equation}
Here, the quark f\mbox{}ields $q$ have colour $(N_c=3)$ and f\mbox{}lavour
indices which are suppressed. We suppose that four-, six-, and eight-quark
interactions ${\cal L}_{4q}$, ${\cal L}_{6q}$, ${\cal L}_{8q}$ are
ef\mbox{}fectively local, for it is known that meson physics in the large
$N_c$ limit is described by a local Lagrangian of this type
\cite{Witten:1979}.
The interaction Lagrangians ${\cal L}_{4q}$ and ${\cal L}_{6q}$ of the model
in the scalar and pseudoscalar channels is given by two terms
\begin{eqnarray}
\label{L4q}
  {\cal L}_{4q} &\!\!\! =\!\!\!& \frac{G}{2}\left[(\bar{q}
  \lambda_aq)^2+ (\bar{q}i\gamma_5\lambda_aq)^2\right], \\
\label{Ldet}
  {\cal L}_{6q} &\!\!\! =\!\!\! &\kappa (\mbox{det}\ \bar{q}P_Lq
                      +\mbox{det}\ \bar{q}P_Rq).
\end{eqnarray}
The f\mbox{}irst one is the $U_L(3)\times U_R(3)$ chiral symmetric
interaction specifying the local part of the ef\mbox{}fective
four-quark Lagrangian in channels with quantum numbers $J^P=0^+, 0^-$.
The Gell-Mann f\mbox{}lavour matrices $\lambda_a,\ a=0,1,\ldots ,8,$
are normalized such that $\mbox{tr} (\lambda_a \lambda_b) =2\delta_{ab}$.
The second term represents the 't Hooft determinantal interactions
\cite{Hooft:1978}. The matrices $P_{L,R}=(1\mp\gamma_5)/2$ are
projectors and the determinant is over f\mbox{}lavour indices. The
determinantal interaction breaks explicitly the axial $U_{A}(1)$
symmetry and Zweig's rule. The global chiral $SU(3)_L\times SU(3)_R$
symmetry of the Lagrangian (\ref{efflag}) at $m=0$ is spontaneously broken
to the $SU(3)$ group, showing the dynamical instability of the fully
symmetric solutions of the theory. In addition, the current quark mass
$m$, being a diagonal matrix in f\mbox{}lavour space with elements
$\mbox{diag} (m_u, m_d, m_s)$, explicitly breaks this symmetry down,
retaining only the reduced $SU(2)_I\times U(1)_Y$ symmetries of isospin
and hypercharge conservation, if $m_u = m_d \neq m_s$.

The eight-quark Lagrangian which describes the spin zero interactions
contains two terms: ${\cal L}_{8q}={\cal L}_{8q}^{(1)} +
{\cal L}_{8q}^{(2)}$ \cite{Osipov:2006}, where
\begin{eqnarray}
   {\cal L}_{8q}^{(1)}&\!\!\! =\!\!\! &
   8g_1\left[ (\bar q_iP_Rq_m)(\bar q_mP_Lq_i) \right]^2,
   \\
   {\cal L}_{8q}^{(2)}&\!\!\! =\!\!\!&
   16 g_2\left[ (\bar q_iP_Rq_m)(\bar q_mP_Lq_j)
   (\bar q_jP_Rq_k)(\bar q_kP_Lq_i) \right].
\end{eqnarray}
Here the sum is taken over f\mbox{}lavour indices $i,j =1,2,3$;
${\cal L}_{8q}$ is a $U(3)_L\times U(3)_R$ symmetric interaction with
OZI-violating ef\mbox{}fects in ${\cal L}_{8q}^{(1)}$. The f\mbox{}irst
term ${\cal L}_{8q}^{(1)}$ coincides with the OZI-violating eight-quark
interactions considered in \cite{Alkofer:1989}. The second term
${\cal L}_{8q}^{(2)}$ represents interactions without violation of Zweig's
rule. ${\cal L}_{8q}$ is the most general Lagrangian which describes the
spin zero eight-quark interactions without derivatives. It is the lowest
order term in number of quark f\mbox{}ields which is relevant to the case.
We restrict our consideration to these forces, because in the long
wavelength limit the higher dimensional operators are suppressed.

We view the main role of eight-quark forces considered as folows:

\noindent
(i) They are of vital importance for the stability of the ground state
built from four and six-quark interactions: the quark model considered
follows the general trend of spontaneous breakdown of chiral symmetry 
and possesses a globally stable ground state, when relevant inequalities
in terms of the coupling constants hold, $ g_1>0, g_1+3g_2>0,
Gg_1>(\kappa /16)^2$ \cite{Osipov:2006}.

\noindent
(ii) The low lying scalar and pseudoscalar mesonic spectra are almost
insensitive to the eight-quark forces \cite{Osipov:2007a}.

\noindent
(iii) The $8q$-interactions play an important role in determining the
critical temperature, $T_c$, at which transitions occur from the dynamically
broken chiral phase to the symmetric phase, lowering the value of $T_c$ with
growing strength of the $8q$ couplings \cite{Osipov:2007c}.

\noindent
(iv) The multi-quark interactions introduce new additional features to
the catalysis of dynamical symmetry breaking by a constant magnetic
f\mbox{}ield $H$ in $3+1$ dimensions: the f\mbox{}irst minimum catalyzed
by a constant magnetic f\mbox{}ield (that is, a slowly varying f\mbox{}ield)
is then smoothed out with increasing $H$ at the characteristic scale $H\sim
10^{19}$G. The reason is that multi-quark forces generate independently
another local minimum associated with a larger dynamical fermion mass. This
state may exist even for multi-quark interactions with a subcritical set of
couplings and is globally stable with respect to a further increase of the
magnetic f\mbox{}ield \cite{Osipov:2007b}.

\noindent
(v) The OZI-violating terms with coupling strength $g_1$ af\mbox{}fect the
mechanism of $\chi SB$: starting from some critical value of the coupling
$g_1=(g_1)_{crit}$ the $\chi SB$ is induced by the $6q$ 't\~Hooft interactions,
as opposed to the $4q$ NJL forces at $g_1<(g_1)_{crit}$ \cite{Osipov:2007a}.

\noindent
(vi) It turns out that the mesonic spectra built on the spontaneously
broken vacuum induced by the 't Hooft interaction strength, as opposed to
the commonly considered case driven by the four-quark coupling, undergo a
rapid crossover to the unbroken phase with a slope and at a temperature
which is regulated by the strength of the OZI violating eight-quark
interactions. This strength $g_1$ can be adjusted in consonance with the
four-quark coupling $G$ (keeping the remaining model parameters f\mbox{}ixed) 
and leaves the spectra unchanged, except for the sigma meson mass which
decreases. A f\mbox{}irst order transition behavior is also a possible
solution within the present approach at large $g_1$ \cite{Osipov:2008}.

\noindent
(vii) They may be also of importance in decays and scattering, not
considered so far.
\vspace{0.5cm}

We are very grateful to the organizers of the "Mini-Workshop Bled 2009:
Problems in multi-quark states", for the kind invitation to present this work.
Research supported by Funda\ca o para a Ci\^encia e Tecnologia, grants FEDER,
OE, POCI 2010, CERN/FP/83510/2008, and by the European Community-Research
Infrastructure Integrating Activity Study of Strongly Interacting Matter
(Grant Agreement 227431) under the Seventh Framework Programme of the EU.


\begin{thebibliography}{99}
\bibitem{Nambu:1961} Y. Nambu, J. Jona-Lasinio, Phys. Rev. 122 (1961) 345;
      124 (1961) 246.
\bibitem{Eguchi:1976} T. Eguchi, Phys. Rev. D 14 (1976) 2755;
      K. Kikkawa, Progr. Theor. Phys. 56 (1976) 947.
\bibitem{Volkov:1982} M.K. Volkov and D. Ebert, Sov. J. Nucl. Phys.
      36 (1982) 736;
      D. Ebert and M.K. Volkov Z. Phys. C 16 (1983) 205.
\bibitem{Ebert:1986}
      M.K. Volkov, Ann. Phys. (N.Y.) 157 (1984) 282;
      A. Dhar and S. Wadia, Phys. Rev. Lett. 52 (1984) 959;
      A. Dhar, R. Shankar and S. Wadia, Phys. Rev. D31 (1985) 3256;
      D. Ebert and H. Reinhardt, Nucl. Phys. B271 (1986) 188.
\bibitem{Bijnens:1993} J. Bijnens, C. Bruno, E. de Rafael, Nucl. Phys. B
      390 (1993) 501.
\bibitem{Simonov:1997} Yu.A. Simonov, Phys. Lett. B 412 (1977) 371;
      Yu.A. Simonov, Phys. Rev. D 65 (2002) 094018.
\bibitem{Hooft:1978} G. 't Hooft, Phys. Rev. D 14 (1976) 3432;
      Erratum: {\it ibid} D 18 (1978) 2199.
\bibitem{Andrianov:1993} A.A. Andrianov, and V.A. Andrianov,
      Int.J.Mod.Phys. A 8 (1993) No.11, 1981.
\bibitem{Bernard:1988} V. Bernard, R.L. Jaffe, U.-G. Meissner, Phys. Lett.
      B 198 (1987) 92; Nucl. Phys. B 308 (1988) 753.
\bibitem{Reinhardt:1988} H. Reinhardt, R. Alkofer, Phys. Lett. B 207
      (1988) 482.
\bibitem{Weise:1990} S. Klimt, M. Lutz, U. Vogl, W. Weise, Nucl. Phys. A
      516 (1990) 429; U. Vogl, M. Lutz, S. Klimt, W. Weise, Nucl. Phys. A
      516 (1990) 469; U. Vogl, W. Weise, Progr. Part. Nucl. Phys. 27 (1991)
      195.
\bibitem{Takizawa:1990} M. Takizawa, K. Tsushima, Y. Kohyama, K. Kubodera,
      Nucl. Phys. A 507 (1990) 611.
\bibitem{Klevansky:1992} S.P. Klevansky, Rev. Mod. Phys. 64 (1992) 649.
\bibitem{Hatsuda:1994} T. Hatsuda, T. Kunihiro, Phys. Rep. 247 (1994) 221.
\bibitem{Witten:1979} E. Witten, Nucl. Phys. B 160 (1979) 57.
\bibitem{Osipov:2006} A.A. Osipov, B. Hiller, J. da Provid\^ encia,
      Phys. Lett. B 634 (2006) 48.
\bibitem{Alkofer:1989} R. Alkofer and I. Zahed, Mod. Phys. Lett. A 4
      (1989) 1737; R. Alkofer and I. Zahed, Phys. Lett. B 238 (1990) 149.
\bibitem{Osipov:2007a} A.A. Osipov, B. Hiller, A.H. Blin,
      J. da Provid\^ encia, Ann. Phys. 322 (2007) 2021.
\bibitem{Osipov:2007c} A.A. Osipov, B. Hiller, J. Moreira, A.H. Blin,
      J. da Provid\^ encia, Phys. Lett. B 646 (2007) 91.
\bibitem{Osipov:2007b} A.A. Osipov, B. Hiller, A.H. Blin, J. da
      Provid\^ encia, Phys. Lett. B 650 (2007) 262; B. Hiller, A.A. Osipov, A.H. Blin, J. da
      Provid\^ encia, SIGMA 4 (2008) 024. 
\bibitem{Osipov:2008} A.A. Osipov, B. Hiller, J. Moreira, A.H. Blin,
      Phys. Lett. B 659 (2008) 270.
\end{thebibliography}
\end{document}